\documentclass[aps,prd,twocolumn,superscriptaddress,showpacs]{revtex4}

\newcommand{\mnras}{Mon.\ Not.\ R. A. S.}
\newcommand{\apjl}{Astrophys.\ J.\ Let.\ }
\newcommand{\apjs}{Astrophys.\ J.\ Supp.\ }
\newcommand{\aap}{Astron.\ \& Astrophys.\ }
\newcommand{\aaps}{Astron.\ \& Astrophys.\ Supp.\ }
\newcommand{\aj}{Astron. J.}
\usepackage{graphicx,bm}

\begin{document}

\title{Constraint on the Post-Newtonian Parameter \bm{$\gamma$} on Galactic Size Scales}

\author{Adam S. Bolton}
\email{abolton@cfa.harvard.edu}
\affiliation{Harvard-Smithsonian Center for Astrophysics, 60 Garden St., Cambridge, MA 02138 USA}
\author{Saul Rappaport}
\author{Scott Burles}
\affiliation{Department of Physics and Kavli Institute for Astrophysics and Space Research, Massachusetts Institute of Technology, 77 Massachusetts Ave., Cambridge, MA 02139 USA}

\date{2006 August 24.  To appear in Phys.\ Rev.\ D15 Rapid Communications.}

\begin{abstract}
We constrain the post-Newtonian gravity parameter $\gamma$
on kiloparsec scales by comparing the masses of 15 elliptical
lensing galaxies from the Sloan Lens ACS Survey
as determined in two independent ways.
The first method assumes only that Newtonian gravity is correct
and is independent of $\gamma$,
while the second uses gravitational lensing which depends
on $\gamma$.  More specifically, we combine Einstein radii
and radial surface-brightness gradient measurements of
the lens galaxies with empirical distributions for the mass
concentration and velocity anisotropy of elliptical galaxies
in the local universe to predict $\gamma$-dependent
probability distributions for the lens-galaxy velocity
dispersions.  By comparing with observed velocity dispersions,
we derive a maximum-likelihood value of
$\gamma = 0.98 \pm 0.07$ (68\% confidence).
This result is in excellent agreement with the prediction of
general relativity that has previously been verified to this
accuracy only on solar-system length scales.
\end{abstract}


\maketitle

Gravitational lenses provide some of the most inspirational and
thought-provoking images in astronomy.  It is often said in the popular press
that these represent a wonderful verification of Einstein's theory of general
relativity (GR).  In 1937, Zwicky \cite{zwicky_1937} proposed that lensing by
distant galaxies and clusters would furnish both a test of GR and a tool for the
measurement of the lensing masses.  Since the discovery of the first gravitational
lenses, however, astronomers have emphasized the latter application over the former.
In fact, gravitational lensing alone cannot separately determine the masses of
lenses and test the weak-field limit of the Schwarzschild metric which underlies
the theory of lensing.

Since Eddington's original solar eclipse expedition of 1919,
the Schwarzschild metric has been extensively probed in the weak field limit within
the solar system~(e.g.~\cite{bertotti_2003}) and with binary radio pulsars
(e.g.~\cite{weisberg_taylor_1984}).  In all cases, however,
the scales involved are of order light seconds.  Most recently, the Cassini
mission has determined the post-Newtonian parameter
$\gamma$~(e.g.~\cite{will_nord_72, mtw}),
described below, by directly measuring the Shapiro delay~\cite{shapiro_1964} as
radio signals pass by the Sun in their travel from the spacecraft to the Earth,
giving a measured value of $\gamma = 1 + (2.1 \pm 2.3) \times 10^{-5}$~\cite{bertotti_2003}.
In this paper we set precise constraints on $\gamma$ on galactic scale sizes of several
kiloparsecs
(1~parsec $\simeq 3.1 \times 10^{18}$~cm):
impact parameters $\sim10^{11}$ times larger than in Sun-grazing
solar-system tests.  We accomplish this by
comparing the masses of relatively low-redshift
($0.06 < z < 0.33$) galaxies as deduced from strong gravitational lensing with
their masses as estimated from their
stellar orbital dynamics.
Since the
former determination depends upon GR through the value of $\gamma$, whereas the
latter depends only upon Newtonian gravity, the quantitative agreement between
the two methods can place a direct constraint upon $\gamma$.
Our analysis constrains $\gamma$ to have a value
of $0.98 \pm 0.07$ (68\% confidence),
in excellent agreement with the GR prediction of $\gamma = 1$.
This statistical precision is enabled by data for a
large and homogeneous sample of recently discovered
gravitational lens galaxies.
Similar analyses have been carried out previously, but with
much lower statistical significance.
Nottale~\cite{nottale_1998} applied this type of test to
the lensing galaxy cluster A370 to obtain $0.87 < \gamma < 1.55$,
subject to assumptions about the mass structure of the cluster.
Dar~\cite{dar_1992} found agreement within $\alt$30\% error bars
between the observed and predicted image separations in a heterogeneous
sample of 5 gravitational-lens systems,
also for fixed mass-structure assumptions,
but did not translate this result
into a constraint on $\gamma$.
\mbox{Sirousse-Zia}~\cite{sirousse_zia_1998} developed formalism
for the test, expressing the dependence upon $\gamma$ of the lensing properties
of the singular isothermal sphere galaxy model (see below),
but did not derive any observational results.

The dependence of gravitational lensing upon $\gamma$
can be derived from a general form of
the metric for a point mass $m$ in the weak field limit for a space-time that
yields Newtonian gravity:
\begin{eqnarray}
d\tau^2 = dt^2 \left(1-\frac{2m}{r}\right) -dr^2
\left(1+\frac{2\gamma m}{r}\right) - r^2 d\phi^2~~.
\end{eqnarray}
The parameter $\gamma$ equals unity for the Schwarzschild metric, $\phi$ is
the angle in the invariant orbital plane, and $G$ and $c$ have been set equal to 1.
In this formulation, the weak-field gravitational acceleration can be calculated,
and is verified to be $-Gm/r^2$ as given by Newton and independent of the parameter
$\gamma$.  From the speed of light in the radial direction as inferred by an external
observer at infinity, one may define an effective index of refraction for the space
surrounding the point mass, and hence around any arbitrary mass distribution through
the principle of superposition.  This quantity is used to compute the Shapiro delay,
and for sufficiently isolated gravitational-lens systems one may invoke the thin-lens
approximation.  Thus gravitational lensing can be formulated in terms of two-dimensional
Fermat time delay surfaces~\cite{schneider_fermat, blandford_narayan}.  By extremizing
the sum of the Shapiro and geometric time delays, one obtains the ``lens equation''
that relates position in the observed image plane to location in the unobserved and
unlensed source plane:
\begin{equation}
\vec{\theta_s} = \vec{\theta} - \frac{(1+\gamma)}{2} \vec{\nabla} \psi(\vec{\theta})~~.
\end{equation}
Here, $\vec{\theta}_s$ is the angular source location, $\vec{\theta}$ is the angular
location of the image, and $\psi(\vec{\theta})$ is a scaled line-of-sight integral
of the Newtonian gravitational potential of the lensing object
(see Eq.~48 of~\cite{nb_1996} for the explicit
definition of $\psi$ in this context).  The Einstein radius,
defined by $\vec{\theta_s} = 0$ for circularly-symmetric projected potentials, simply
scales with the factor $(1+\gamma)/2$.  The difficulty in inferring anything about
the $\gamma$ parameter is that it appears only as a product with the relevant masses
of the problem.  Unless one knows the lensing mass via some method other than lensing,
there is no distinction between, e.g., the solution for a
lensing mass $M$ and $\gamma = 1$
and the solution for a lensing mass of $2M$ and $\gamma = 0$.

The actual observational test that we perform here
can be understood most simply in terms of the
singular isothermal sphere (SIS) galaxy model,
though we in fact allow for more general galaxy models
in our analysis.
The SIS is a dynamically self-consistent, spherically
symmetric three-dimensional profile with
density $\rho(r) = \sigma_v^2 / (2 \pi G r^2)$.  It is perhaps the best one-parameter
model for elliptical galaxies, characterized by an isotropic
and radially constant dispersion of stellar
orbital velocities $\sigma_v$ that determines the mass within any given
radius.  Through gravitational lensing, the SIS forms a
ring image of any background objects along the same line of sight,
with an angular ``Einstein radius'' of
\begin{equation}
\label{theta_e}
\theta_E = (1 + \gamma) 2 \pi (\sigma_v^2 / c^2) (D_{LS} / D_S)~~,
\end{equation}
where $D_{LS}$ and $D_S$ are distance measures discussed further below.
Thus, within the context of the SIS model, measurements of the
Einstein radii {\em and} velocity dispersions of gravitational
lens galaxies can be used to constrain the
$\gamma$ parameter~\cite{sirousse_zia_1998}.
(From an observational standpoint,
velocity dispersion is defined in this work as the luminosity-weighted
integral of the second moment of the stellar velocity distribution
along the observational line of sight.  It may be measured from the spectrum
of a galaxy whose stellar population is spatially unresolved
by fitting for the broadening of stellar atomic
absorption lines that best reproduces
the features seen in the integrated galaxy spectrum.)

Spatially resolved, high signal-to-noise observations
of the kinematics of elliptical galaxies, whose
stellar populations are dynamically ``hot''
(i.e., not characterized by ordered circular orbits),
can be used in combination with detailed modeling
to deduce radial density profiles and to directly
test the validity of the isothermal approximation.  Such studies
(e.g.\ \cite{vandermarel_1991, vdm_franx_1993, kronawitter_2000, gerhard_2001,
romanowsky_kochanek, emsellem_2004, cappellari_2006}) indicate
a modest amount of dark matter within one half-light radius
and an approximately isothermal density profile
in elliptical galaxies that are nearby enough to permit the
necessary observations.
Gravitational lens galaxies (hereafter GLGs), by contrast,
are typically elliptical galaxies at relatively high
redshift and with significant contamination from the light of lensed quasar
images, and thus do not allow the same type of dynamical analysis as local
ellipticals.  Thus lensing itself is usually the only robust measurement of the
mass of the GLG, and a model for the density profile
must simply be assumed.
An exception is the Lenses Structure and
Dynamics [LSD] Survey \cite{kt02, kt03, tk02, tk03, tk04}, which
has used lens-galaxy velocity dispersions in combination
with Einstein radii and the implicit GR assumption of $\gamma=1$ to place
constraints on the relative contributions of luminous and dark matter
in distant GLGs.

The current work is enabled by the
significant new sample of relatively low redshift
($0.06 < z < 0.33$) elliptical GLGs
presented by the Sloan Lens ACS (SLACS)
Survey \cite{slacs1, slacs2, slacs3}.
The SLACS GLGs,
which were discovered within the Sloan Digital Sky Survey
spectroscopic database
(SDSS:~\cite{york_sdss}), are
particularly distinguished by the ease and accuracy with which their surface
brightness profiles and stellar dynamics can be measured, relative
to previously known GLGs.
Publicly available SDSS velocity-dispersion measurements of
SLACS GLGs provide a single, lensing-independent dynamical mass scale
for each system,
though they do not allow a lensing-independent determination of the
GLG density profiles.
In this work we adopt the hypothesis that, by virtue of their
relatively low redshifts, SLACS GLGs are sufficiently like
nearby elliptical galaxies for their \textit{distribution} in density
profiles to be approximated by the
distribution deduced for nearby elliptical galaxies from the application
of detailed dynamical modeling.  The masses of SLACS GLGs can thus
be inferred independently of lensing from their SDSS
velocity dispersions, with an uncertainty quantified
by both measurement error and intrinsic scatter in
density profile (and velocity anisotropy, discussed below).
The most significant evolution to be expected
between nearby elliptical galaxies and the SLACS GLGs
is in luminosity, which we circumvent by working
entirely with mass, shape, and dynamical observables.

In order to work with directly observable quantities
(of which galaxy mass is not one),
we frame our analysis in terms of a comparison between
the stellar velocity dispersions of GLGs (i) as observed
and (ii) as predicted from their lensing
Einstein radii for a given value of $\gamma$.
Before proceeding into more detail, we note that the
observed stellar dispersions of SLACS GLGs agree
within observational errors
with their Einstein radii when the latter are directly translated into velocity
dispersions
using Eq.~\ref{theta_e}
and assuming
$\gamma=1$ (see Fig.\ 5 of~\cite{slacs2}).
To incorporate the possible effects of non-isothermal lens
profiles, we employ the self-similar,
axisymmetric model described by Koopmans~\cite{lvek_lensdyn}.
This model approximates the mass- and luminosity-density profiles of
GLGs with power-law forms: i.e.\ three-dimensional
mass density $\rho (r) \propto r^{-\alpha}$ and luminosity density
$\nu (r) \propto r^{-\delta}$ (here we have used $\alpha$ to replace the
$\gamma^{\prime}$ of~\cite{lvek_lensdyn} to avoid confusion with the
post-Newtonian parameter of interest).
With $\alpha$ and $\delta$
considered as separate parameters, the model corresponds to a scale-free
galaxy with a constant logarithmic radial mass-to-light ratio gradient,
capable of describing an increasing dark-matter fraction at
increasing radius.
Larger values of $\alpha$ and $\delta$
correspond to more centrally concentrated mass and light profiles.
The observed angular Einstein radius $\theta_E$ of a GLG gives
a measurement of the enclosed mass.
Taking the mass normalization set by $\theta_E$ and fixed
values for $\alpha$ and $\delta$, the spherically-symmetric
steady-state Jeans
equation is integrated analytically for velocity dispersion in
the radial direction as a function of radius.  (The Jeans equation
is obtained from the velocity moments of the collisionless Boltzmann
equation for the phase space distribution of stellar orbits in
Newtonian gravity; see, e.g., \cite{bt})
The model further assumes a
constant radial profile for the conventionally defined velocity anisotropy
parameter $\beta$
which relates velocity dispersions in the radial and
tangential directions:
$\beta = 1 - \sigma^2_{v,\mathrm{tan}} / \sigma^2_{v,\mathrm{rad}}$.
The Jeans-equation solution assumes
the validity of Newtonian gravitation on the relevant scales,
but is independent of $\gamma$
in the weak-field limit as discussed above.
The luminosity-weighted squared velocity dispersion is integrated
along the line of sight to give an analytic prediction for the
observed velocity dispersion as a function of position within
the image of the GLG.
The simplest SIS model is described by the special case of
$\alpha = \delta = 2$ and $\beta = 0$.

The sample we analyze consists of the 15 SLACS GLGs with published
Einstein radii, determined by fitting singular isothermal
ellipsoid (SIE) lens models with the
normalization of~\cite{kormann_sie} to \textsl{Hubble Space Telescope}
(\textsl{HST}) imaging data~\cite{slacs3}.
Though the SIE model assumes a particular radial mass
profile, measured Einstein radii depend only weakly upon the details of the
radial mass distribution of the GLG.  We obtain $\delta$ values by computing
average logarithmic surface-brightness profile slopes
for the 15 systems through non-linear least-squares fitting of elliptical
power-law luminosity models convolved with the instrumental point-spread function
to the \textsl{HST} imaging data within a radius of 1.8 arcseconds
about the center of each GLG.
(A three-dimensional brightness profile $\nu (r) \propto r^{-\delta}$
will have a projected two-dimensional profile $I (R) \propto R^{-\delta + 1}$.)
Both the mass and light models fitted to the \textsl{HST} data include
a projected
major-to-minor
axis-ratio parameter $q$ to allow for ellipticity.
To connect these models to the axisymmetric approximation
of the analytic Jeans equation-based framework, we use the interchange
\begin{equation}
R \leftrightarrow R_q = \sqrt{qx^2 + y^2 / q} ~~,
\end{equation}
which conserves the total mass or light within a given
iso-density or iso-brightness contour.
Stellar velocity dispersion
values for each GLG are taken from the
output of the Princeton 1d spectroscopic
pipeline analysis of
the SDSS database~\cite{schlegel_1d}.
The SDSS spectrograph integrates all atmospherically
blurred galaxy light within a circular optical-fiber aperture of
radius 1.5 arcseconds on the sky.
The fractional systematic velocity-dispersion inaccuracy incurred in
a circularly-symmetric Jeans analysis of an oblate
galaxy is calculated by~\cite{kochanek_dyn_halo_94}
to be small provided that one uses angularly-averaged velocity
dispersions and Einstein radii,
as we do in our calculations here.  We add this error
in quadrature to the observational velocity-dispersion
error estimates for each lens, though it is small (median contribution
of 2\%) and has a negligible effect on our result.

To determine suitable probability distributions
for the mass concentration $\alpha$
and velocity anisotropy $\beta$ of the SLACS GLGs, we make use of the
results of Gerhard et al.~\cite{gerhard_2001},
who compute the density structure
for a sample of nearby elliptical galaxies using dynamical observations.
From Fig.~1 of that work, we determine the change in circular velocity
(or, equivalently, mass enclosed)
between 0.2 and 0.6 half-light radii for each of the 17 galaxies
with data over that range.  The chosen radial range corresponds
roughly to the range probed by the SLACS GLG sample.
We assign to each of these galaxies a logarithmic radial density slope $\alpha$
that will give the same relative change in
circular velocity.  This gives a sample mean log-slope of
$\langle \alpha \rangle = 1.93$ and an intrinsic RMS variation of
$\sigma_{\alpha} = 0.08$.  From Fig.~5 of the same work, we
calculate the average and RMS variation of the
velocity anisotropy
parameter $\beta$ (e.g.~\cite{bt}) to be $\langle \beta \rangle = 0.18$ and
$\sigma_{\beta} = 0.13$ (excluding the outlier galaxies
NGC 4636 and NGC 4486B).  We model the probability distributions
of $\alpha$ and $\beta$ for the SLACS GLGs as uncorrelated
Gaussians with these parameters.  Neither of these properties exhibit
significant correlation with galaxy mass in~\cite{gerhard_2001}.

For comparison with the observed velocity dispersions, we integrate
the position-dependent velocity dispersion predicted by lensing over the
SDSS spectroscopic
optical-fiber
aperture.  The
combined effect
of the circular
SDSS aperture (radius 1.5 arcseconds) and the typical
image quality
of FWHM 2 arcseconds
(due to atmospheric blurring)
is well approximated by a circular
Gaussian weighted aperture with a FWHM of
$2.8$ arcseconds,
which we combine with luminosity weighting in an integration of
Eq.~2.4 of~\cite{lvek_lensdyn}.  We furthermore incorporate the
dependence of gravitational lensing upon $\gamma$
and eliminate the explicit dependence on the mass of the GLG
in favor of the Einstein radius to obtain
\begin{eqnarray}
\label{eq:vdisp}
\nonumber
\langle \sigma_{\mathrm{l.o.s.}}^2 \rangle = {{c^2} \over {4 \pi}}
{{D_S} \over {D_{LS}}} \theta_E \left( {2 \over {1 + \gamma}} \right)
f(\alpha, \delta, \beta) \left( {{\theta_A} \over {\theta_E}} \right)^{2 - \alpha} \\
\times 2^{1-\alpha/2} \left( {{5 - \delta - \alpha} \over {3 - \delta}}\right)
\left( {{\Gamma [(5-\delta-\alpha)/2]} \over {\Gamma [(3-\delta)/2]}} \right) ~~.
\end{eqnarray}
Here, the first occurrence of $\theta_E$ is to be expressed in radians.
The dimensionless function $f(\alpha, \delta, \beta)$ is given
in~\cite{lvek_lensdyn}.
The quantity $\theta_A$ is the ``Gaussian sigma'' of the
spatial weighting aperture: i.e., $2.8$ arcseconds divided by $2.355$.
Factors on the second line arise from our use of a Gaussian
integration aperture.
$D_S$ and $D_{LS}$ are angular-diameter
distances from the observer to the lensed galaxy and from the GLG
to the lensed galaxy respectively,
specifying the angle subtended by a given physical
distance transverse to the line of sight.
These distances are functions of the
redshifts of the two galaxies, which have been measured
to high accuracy for all SLACS systems using SDSS spectroscopic data~\cite{slacs1}.
$D_S$ and $D_{LS}$ also depend upon the
adopted cosmology, and in principle would be altered through the Robertson-Walker
metric by any change in $\gamma$
(as noted by~\cite{sirousse_zia_1998}).  However, such changes would necessarily
be accompanied by changes in the matter-energy density of the universe in order
to reproduce the shape of the luminosity distance-redshift relation empirically
constrained by type Ia supernova observations out to redshift
$z\approx 1$~\cite{riess_1998, perlmutter_1999, knop_2003, riess_2004}.
Thus although we shall compute
$D_S$ and $D_{LS}$ under the assumption of the currently favored
cosmology with density parameters
$(\Omega_M, \Omega_\Lambda) = (0.3,0.7)$~\cite{spergel_wmap}, we regard this
recipe as a suitable proxy for the empirical distance-redshift relation.

We use Eq.~\ref{eq:vdisp} to determine numerically the probability
distribution of the ``true'' aperture velocity dispersion
$\langle \sigma_{\mathrm{l.o.s.}}^2 \rangle^{1/2}$, which
we denote below by $\sigma$ for convenience, for each
GLG as a function of $\gamma$.
This is done by computing $\sigma$ over a grid of all physically
allowed $\alpha$ and $\beta$ values (i.e., those that give finite
central mass and predict non-negative $\sigma^2$) at fixed
$(\delta, \theta_E, \theta_A)$, assigning differential weights
over the grid as given by the assumed probabilities of
$\alpha$ and $\beta$, and sorting in $\sigma$ to
determine the cumulative probability distribution.
This computation is repeated over a grid of $\gamma$
values from 0--4 to derive the desired
probability density $P_1 (\sigma | \gamma)$ for each lens.
The probability density for obtaining a particular
\textit{measured} SDSS velocity dispersion $\sigma_{\mathrm{SDSS}}$
at fixed $\gamma$ is then given by
\begin{equation}
\label{eq:psigma}
P_3 (\sigma_{\mathrm{SDSS}} | \gamma) = \int d\sigma \,
P_2 (\sigma_{\mathrm{SDSS}} | \sigma) P_1(\sigma | \gamma) ~~.
\end{equation}
For $P_2(\sigma_{\mathrm{SDSS}} | \sigma)$, we
assume a Gaussian distribution about the true value,
with a width given by the measurement error
for each system.  Statistical errors on the measured $\delta$
and $\theta_E$ values are neglected, since
they have an exceedingly small effect
compared to the errors on the $\sigma_{\mathrm{SDSS}}$ values
and the assumed intrinsic variation in $\alpha$ and $\beta$.

\begin{figure}[t]
\centerline{\scalebox{0.4}{\includegraphics{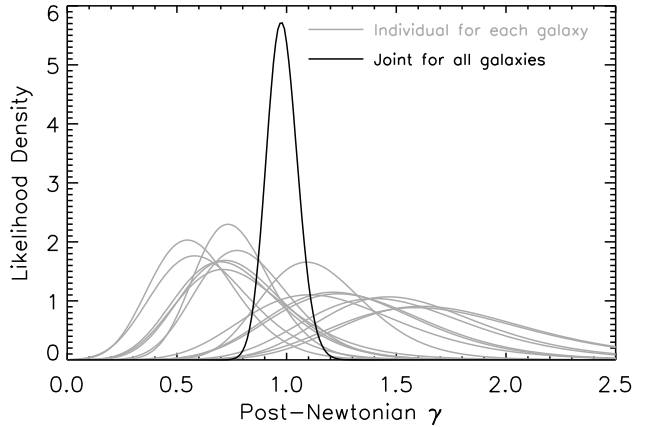}}}
\caption{
\label{fig:gamma}
Normalized likelihood densities for the post-Newtonian parameter
$\gamma$ derived from individual SLACS GLGs
(grey), with joint likelihood for the entire
sample (black).  The joint distribution
is approximately Gaussian,
giving $\gamma = 0.98 \pm 0.07$ (68\% confidence).}
\end{figure}

For each GLG $i$, given its measured $\sigma_{\mathrm{SDSS}}$,
Eq.~\ref{eq:psigma} may be interpreted as a likelihood
distribution for the parameter $\gamma$.
Since we expect the true value of $\gamma$ to be universal
and thus the same for all GLGs in the sample,
we can express its likelihood distribution given all the
data and the prior assumptions as
\begin{equation}
\mathcal{L} (\gamma) = \prod_{\mathrm{lenses}~i}
P_3^{(i)} (\sigma_{\mathrm{SDSS}}^{(i)} | \gamma) ~~.
\end{equation}
Fig.~\ref{fig:gamma} shows the individual factors
in this likelihood product, as well as the joint
normalized likelihood density.
Though the distribution of individual maximum-likelihood
values seen in Fig.~\ref{fig:gamma} appears slightly
bimodal between $\gamma$ values greater than 1 and less than 1,
this effect is not statistically significant:
a Kolmogorov-Smirnov test of the distribution of individual
$\gamma$ values indicates a reasonable 33\% probability of being drawn
from a Gaussian distribution about $\gamma=1$ with a width
equal to the mean RMS width of the individual $\gamma$
distributions (approximately $0.3$).  Furthermore, the individual $\gamma$
values are not significantly
correlated with any lens-galaxy observables.
The joint distribution $\mathcal{L} (\gamma)$
is described well by a Gaussian in $\gamma$,
from which we derive a maximum-likelihood constraint of
$\gamma = 0.98 \pm 0.07$ (68\% confidence).

Our maximum-likelihood value for $\gamma$ is in excellent
agreement with the GR-predicted value of $\gamma=1$.
This result is an important quantitative test of the theory
of gravitation
on scales
much larger than have been probed previously with
solar-system experiments.  The current analysis is enabled by
the new sample of GLGs from the SLACS
Survey for which lens masses can (implicitly) be estimated
through a combination of observables and reasonable
assumptions independent of lensing itself.
We note that a spatially resolved and/or higher
signal-to-noise observational determination of the
stellar dynamics of the SLACS GLGs would remove
some of the reliance on prior assumptions based
on the local universe, thus making the result more robust.

\begin{acknowledgments}
The authors thank Scott Hughes for valuable feedback
about the manuscript.
ASB thanks Margaret Geller and Virginia Trimble
for questions that first
provoked his thoughts on this subject.
SR acknowledges support from NASA Chandra Grant NAG5-TM5-6003X.
SB acknowledges support from NSF Grant AST-0307705.
Based on observations made with the NASA/ESA Hubble Space Telescope,
obtained from the Data Archive
at the Space Telescope Science Institute,
which is operated by AURA,
Inc., under NASA contract NAS 5-26555.
These observations are associated with program \#10174
(PI: L.V.E. Koopmans).
Funding for the SDSS and SDSS-II has been provided by
the Alfred P. Sloan Foundation, the Participating Institutions,
the National Science Foundation, the U.S. Department of Energy,
NASA, the Japanese Monbukagakusho, the Max Planck Society, and
the Higher Education Funding Council for England.
The SDSS Web Site is \url{http://www.sdss.org/}.
The SDSS is managed by the Astrophysical Research
Consortium for the Participating Institutions. 
\end{acknowledgments}


\end{document}